\documentclass[aps,onecolumn,showpacs, preprintnumbers, amsmath, amssymb, superscriptaddress, aps]{revtex4}
\usepackage{graphicx}
\usepackage{dcolumn}
\usepackage{color}
\usepackage{bm}
\def\Am3{\AA$^{-3}$}
\def\beq{\begin{equation}}
\def\eeq{\end{equation}}
\def\beqa{\begin{eqnarray}}
\def\eeqa{\end{eqnarray}}
\def\beqa{\begin{eqnarray}}
\def\eeqa{\end{eqnarray}}
\def\ri{\mbox{i}}
\def\rd{\mbox{d}}
\def\p{\partial}
\def\re{\mbox{e}}
\def\la{\langle}
\def\ra{\rangle}
\begin{document}

\author{A.B. Kuklov}
\affiliation{Department of Engineering Science and Physics,
CSI, CUNY, Staten Island, NY 10314, USA}

\author{A. M. Tsvelik}
\affiliation{Department of Condensed matter Physics and Materials Science, Brookhaven National Laboratory, Upton, NY 11973, USA}
\title{Parafermion excitations in superfluid of quasi-molecular chains} 


\begin{abstract}
We study a  quantum phase transition in a system of  dipoles confined in
a stack of $N$ identical  one-dimensional lattices (tubes) polarized perpendicularly to the lattices. In this arrangement the intra-lattice interaction is purely repulsive preventing the system collapse and the inter-lattice one is attractive.  The  dipoles
may represent polar molecules or indirect excitons. The transition separates two phases; in one of them superfluidity (understood as algebraic decay of the corresponding correlation functions) takes place in each individual lattice, in the other (chain superfluid) the order parameter is the product of bosonic operators from all lattices.  
We argue that in the presence of finite inter-lattice tunneling the transition belongs to the universality class of 
  the $q=N$ two-dimensional classical Potts model. For $N=2,3,4$ the corresponding low energy field theory is the model of Z$_N$ parafermions perturbed by the thermal operator. Results of Monte Carlo simulations are consistent with these predictions.
The detection scheme for the chain superfluid of indirect excitons is outlined.  
\end{abstract}

\maketitle

\section{Introduction}

Emergence of Majorana fermions (see in \cite{Wilczek}) in topological insulators \cite{TopIn} has inspired a search for such fermions in other condensed matter systems. 
In this article we show that parafermions\cite{expl}, of which Majorana fermions represent a particular case, describe excitation spectra of quantum chains (strings)  of polarized dipoles. Material realization of such systems has became possible due to the recent breakthroughs in creating and trapping 
high density samples of (polar) molecules 
~\cite{Jin-Nagerl}.  As proposed in Ref.\cite{exc}, multi-layered structures of indirect
excitons \cite{Yudson} may also form similar systems in the form of excitonic chains. Each indirect exciton (not to be confused with the excitons formed at non-$\Gamma$ point)
has static dipole moment due to a spatial separation of electron and hole. Interaction between the dipoles in the $N$-layered structure can encourage  a  formation of the excitonic chains similar to chains of polar molecules. Since light field $E$ and excitons are coupled linearly a state of excitonic field $\psi$  is imprinted directly on the emitted light. As a consequence, properties of excitonic chains can be explored through light emission providing a new powerful experimental tool to study strongly correlated systems.  


So far, quantum chains have been studied in various analytical approximations which  neglect tunneling of particles along the chains. In Ref.\cite{Wang} it has been proposed that stiff dipolar non-interacting quantum chains may form Bose-Einstein condensate. Inter-layer pairing in bilayered 2D dipolar fermionic systems  has been studied in the BCS approximation in Refs.\cite{bilayers}. The dimerization transition in the 2D multi-layered geometry of dipolar fermions was analyzed in Ref.\cite{multi2D}, and it has also been proposed that for  strong dipolar interactions long chains can form by  the N-clock phase transition \cite{multi2D}. Fermionic dipolar molecules forming a  mixture of single fermions, dimers and fermionic trimers in 1D $N=3$-layered system has been discussed in the ideal gas approximation in Ref.\cite{ideal1D}.

In low dimensions  quantum fluctuations are enhanced and therefore quantum particles from different 1d tubes may develop strong correlations. This can lead to interesting physics. The difficulty is that such system, in general, is not amenable to the standard mean field or perturbation expansion methods and one has to resort to a combination of non-perturbative techniques and numerics. 
A numerical study of quantum chains which takes into account the partial-chain exchanges as well as the intra-chain dynamics has been performed in Ref.\cite{JLTP} for the case of zero inter-layer tunneling. It has been shown that polar molecules in the $N$-layered geometry can form flexible (quantum rough) chains, and these chains can undergo a quantum phase transition to a superfluid phase characterized by off-diagonal long-range order (ODLRO) in the $N$-body
density matrix, while all $M$-body density matrices with $M<N$ show insulating behavior  regardless of the filling factor (provided it is the same in each layer). If the inter-layer (dipolar) interactions are weak, a stack of $N$ layers features a $N$-component superfluid (N-SF). Once the interaction becomes stronger, the non-dissipative drag between layers will eventually convert the N-SF phase to the flexible chain superfluid (CSF) characterized by ODLRO only in the $N$-body density matrix \cite{JLTP}. The corresponding transition is continuous in the 1D-geometry and discontinuous in the 2D-geometry for $N>2$ \cite{JLTP}. 

\begin{figure}
\centerline{\includegraphics[angle = 0,
width=0.5\columnwidth]{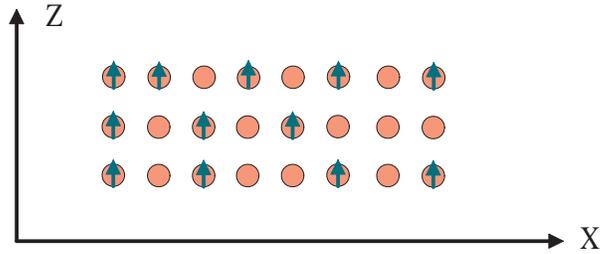}} 
\caption{(Color online)
  Schematic picture  of $N=3$ parallel  1d lattices (also called  "tubes") stretched along the X-axis. The dipole particles (represented by arrows) occupy lattices sites (full circles).  The arrows show the  polarization of the particles  dipole moments (along the Z-axis).  
The dipoles may tunnel between nearest sites along the X and the Z directions.} \label{pict1}
\end{figure}

In the present work we study the transition between N-SF and  CSF states in the presence of finite inter-tube tunneling. Since for 2D-layers  ($N>2$) the transition is of the first order and the inter-layer tunneling cannot change this, we concentrate on the 1D-geometry depicted in Fig.~\ref{pict1}, that is, when layers may  be considered as tubes. 

Our main findings are the following. A quantum phase transition into the $N$-chain superfluid (in 1+1 dimensions) is in the universality of the classical $q=N$ 2D Potts model. That is, for $N=2,3,4$ the transition is a continuous one and for $N>4$ it is of the first order. For $N=2$ we develop a microscopic  low energy  description in terms of  the field theory of two species of Majorana fermions and one gapless bosonic field. For $N=3,4$ instead of a detailed derivation we present arguments based on symmetry of the problem and on results of our numerical calculations.  

 The paper is organized as follows. In Sec.~\ref{phases} we discuss possible phases. Then, in Sec.~\ref{Majorana} we will describe the microscopic lattice model accounting for a system of N coupled tubes. Then we formulate field theoretical description of the system valid in the continuum limit. This description is rigorously derived for the case of two tubes ($N=2$) where the low energy limit leads to a model  of relativistic Majorana fermions. The correlation functions characterizing light emission from the $N=2$ excitonic system are derived. We also present plausible arguments concerning possible field theory for the cases $N=3,4$. These arguments are supported by the Monte Carlo calculations presented in Sec.~\ref{numerics}. The Monte Carlo procedure is performed  for  the coarse-grained dual version of the Hamiltonian in the discretized time approximation. Finally, in the Conclusion we will give a summary of the main results and perspectives for detecting excitonic CSF by $N$-photon correlation spectroscopy.

\section{Order parameters and universality of the transition to the CSF phase}\label{phases}
Without inter-tube tunneling in the N-SF state \cite{JLTP} each tube is characterized by its own phase. 
In the case of finite inter-tube tunneling (along $Z$ in Fig.~\ref{pict1}) these phases lock in into a single phase field $\varphi $ so that the superfluid (SF) is characterized by the bosonic operator $ \psi \sim {\rm e}^{i\varphi }$. 
As is well known,  in 1D the real long range order is substituted by quasi long range order characterized by nonzero stiffness and algebraic decay of certain correlation functions at zero temperature. The SF phase of our system is characterized by an algebraic decay of the bosonic field. Meanwhile in the Chain Superfluid Phase (CSF) correlators of individual Bose operators $\langle \psi^\dagger(x,z) \psi(x',z') \rangle$ decay exponentially with respect to $|x-x'|$ and an algebraic order pertains to the product of operators of all tubes
\beq
\Psi_N (x)= \prod_{z=1,2,...,N}\psi(x,z),
\label{Psi}
\eeq
describing the order of quasi-molecular complexes each consisting of $N$ bosons. 

It is important that  $\Psi_N$, Eq.(\ref{Psi}), is invariant with respect to the transformation 
$ \psi(x,z) \to \exp(2\pi i m(z,x)/N) \psi(x,z)$
where $m(z,x)$ is defined modulo $N$ and obeys the constraint $\sum_z m(z,x) =p N, p=0, 1, 2,... $. 
Thus, $m(z,x) $ can be broken as $m=m'+\tilde{m}$ into the discrete global part $\tilde{m}=p,\,\, p=1,2,...,N-1$ and the local gauge-type $m'(z,x)$ obeying $\sum_z m'=0$. 

Setting aside the discussion of a possible role of the local-gauge symmetry, we note that 
the global transformation forms a discrete symmetry group which determines the universality of the SF-CSF transition. Among the possible candidates one can consider the p-clock model and the (standard) Potts model (see in Ref.\cite{Wu}). While the case $N=2$ should be assigned to the Ising universality class \cite{Sachdev},
the nature of the transition for $N>2$ is not obvious at all. Na\"ively, one may anticipate the p-clock universality because of the nearest neighbor tunneling (between tubes). In what follows we will show that such expectation is not correct, and the criticality is controlled by the standard Potts model (also called as Ashkin-Teller-Potts model). 
Accordingly, for 1D tubes (that is, $D=1+1$ membranes) it should be continuous for $N=2,3,4$ and discontinuous for $N>4$. 

\section{Microscopic Hamiltonan and the effective model for $N=2$ in terms of Majorana fermions \label{Majorana}}
Each tube represents an optical lattice occupied by particles with Bose statistics (polar molecules or indirect excitons).  
The microscopic Hamiltonian $H$ describing SF and CSF has the following form:
\beqa
H = -\sum_{x,z}\Big[t_{||}\Big(a^{\dagger}_{z,x+1}a_{z,x} + h.c.\Big) + t_{\perp}\Big(a^{\dagger}_{z+1,x}a_{z,x} + h.c.\Big)\Big]  +  \frac{1}{2}\sum_{xz; x'z'}V_{xz;x'z'}n_{xz}n_{x'z'},  
\label{Habi}
\eeqa
Here $a^\dagger_{xz},\, a_{xz}$ are creation  (annihilation) operators creating (destroying) a boson  at site $x$ belonging to $z$th tube; 
$n_{xz}=a^\dagger_{xz} a_{xz}$ denotes onsite density operator obeying the hard-core constraint; $V_{xz;x'z'}$ 
describes the matrix element for dipole-dipole interaction between sites $(xz)$ and $ (x'z')$. It is characterized
by the strength $V_d=d_z^2/b^3_z$, where $d_z$ stands for the induced dipole moment and $b_z$ denotes distance between two nearest tubes. This interaction is mainly attractive along the $z$-direction and repulsive along the $x$-direction. In this paper we will be studying a simplified version of the model (\ref{Habi}).
Specifically, we will reduce the dipole-dipole interaction to just a nearest-neighbor attraction $V_1$ along the Z-direction and nearest-neighbor repulsion $V_0$ along the X-direction. Clearly,  such approximation cannot change neither the low energy physics nor  universality class of the transition.  

\subsection{Two tubes ($N=2$). Low energy decsription}
In the low energy limit the microscopic Hamiltonian (\ref{Habi}) can be replaced by the effective model describing the SF to CSF transition (in the chosen approximation). Taking into account the single occupancy constraint we can rewrite Hamiltonian (\ref{Habi})  in terms of the Pauli matrix  operators. Restricting ourselves to the simplest case of two tubes, we have:
\begin{eqnarray}
 H - \mu N&&= 
\sum_j\Big\{\sum_{z=1,2}\Big[ - t_{\parallel}(\sigma^+_{j,z}\sigma^-_{j+1,z} + h.c.) + V_0\sigma^z_{j,z}\sigma^z_{j+1,z} - \mu\sigma^z_{j,z}\Big] \nonumber\\ %
&&  + 
 t_{\perp}(\sigma^+_{j,1}\sigma^-_{j,2} +h.c.) - V_1\sigma^z_{j,1}\sigma^z_{j,2}\Big\} \label{N2}
\end{eqnarray}
where $\sigma^{\pm}$ operators stand, respectively, for the bosonic creation and annihilation ones $a^{\dagger}, a$ in Eq.(\ref{Habi}) and $\hat n = \sigma^z +1$. We assume that here as everywhere throughout the paper the density fluctuations (the total one and the difference between the tubes) are incommensurate with the lattice. In the context of model (\ref{N2}) it is achieved by a proper choice of the chemical potential $\mu$.

Following Schulz \cite{Schulz} we will treat this model at low energies using bosonization technique (see also Ref.\cite{boson}).  
The model describing each  tube is the spin S=1/2 XXZ  model; in the continuous limit  it is equivalent to the Gaussian model:
 \beq
H_a = \frac{v}{2}\int \rd x \Big[K^{-1}(\p_x\Phi_a)^2 + K(\p_x\Theta_a)^2\Big], \label{Gaussian}
\eeq
 where $a=1,2$ labels the tubes and $\Theta_a$ is the field dual to $\Phi_a$: 
 $[\p_x\Theta_a (x), \Phi_a(y)] = -i\delta(x-y)$. The Luttinger parameter $K$ and velocity $v$ are determined by the intra-chain interactions and the chemical potential.  For model (\ref{N2}) at $\mu =0$ we have \cite{johnson}: 
\beqa
K= \frac{\pi}{2 \arccos(V_0/2t_{\parallel})},\quad v= \frac{4 \sqrt{(2t_\parallel)^2 -V^2_0}}{\pi -\arccos(V_0/2t_\parallel)} 
\eeqa
 and we assume that $t_{\parallel} > 0$. 
 Then the continuum limit of the spin operators is given by the following bosonization formulae:
\begin{widetext}
\beqa
&& \sigma^{+}_a(x) = \frac{1}{(2\pi a_0)^{1/2}}\re^{\ri\sqrt{2\pi}\Phi_a} + C\Big[\re^{\ri\sqrt{2\pi}(\Theta_a + \Phi_a) +2\ri k_Fx} + \re^{\ri\sqrt{2\pi}(-\Theta_a + \Phi_a) -2\ri k_Fx}\Big] + ...\nonumber\\
  && \sigma^z_a(x) = \frac{1}{\sqrt\pi}\p_x\Theta_a + \frac{C^z}{(2\pi a_0)^{1/2}}\sin(2k_F x + \sqrt{2\pi}\Theta_a)(-1)^n +... \label{spins}
  \eeqa 
  \end{widetext}
 where dots stand for less relevant operators and $C,C^z$ are amplitudes determined by the short range physics  and $a_0$ is the short range cut-off. 
 The Fermi momentum $k_F$ for each chain is determined by its chemical potential $\mu$ with the convention that $k_F(\mu =0) =0$. As we have mentioned above,  we will always assume that 
 $\mu \neq 0$ so that the spin fluctuations are incommensurate with the lattice.

Substituting (\ref{spins}) into (\ref{N2}) and defining  the fields 
\beqa
&& \Phi_{1,2} =  \Big(K^{1/2}_+\Phi_+ \pm K^{1/2}_-\Phi_-\Big), \nonumber\\
&&\Theta_{1,2} = \frac{1}{2}\Big(K^{-1/2}_+\Theta_+ \pm K^{-1/2}_-\Theta_-\Big),\nonumber\\
&& [\p_x\Theta_a(x),\Phi_b(y)] = -\ri\delta_{ab}\delta(x-y),
\eeqa
 we obtain the Hamiltonian $H = H_+ + H_-$, where $H_+$ describes  the symmetric mode $(+)$: 
\beq
H_+ = \frac{v_+}{2}\int \rd x \Big[(\p_x\Phi_+)^2 + (\p_x\Theta_+)^2\Big], \label{Gaussian2}
\eeq
and $H_-$ contains only the anti-symmetric fields: 
\beqa
 H_- = \int \rd x \Big\{ \frac{v_-}{2}\Big[(\p_x\Phi_-)^2 + (\p_x\Theta_-)^2\Big] + 
\tilde t_{\perp}\cos\Big(\sqrt{4\pi K_-}\Phi_- \Big)- \tilde V_1\cos\Big(\sqrt{4\pi/K_-}\Theta_-\Big)\Big\} \label{charge2},
 \eeqa
 where $\tilde V_1 \sim V_1, \tilde t_{\perp}\sim t_{\perp}$ and 
 \beq
 K_{\pm} = K \pm \frac{V_1}{2\pi v}.
 \eeq

At this juncture we note that the Hamiltonian (\ref{Gaussian2}) describes a mode which is always  critical. The corresponding order parameter is $\Psi_{+} = \sigma^+_1\sigma^+_2$; according to (\ref{spins}) \beq
\Psi_+ \sim \exp(i\sqrt{4\pi K_+}\Phi_+). \label{psi2}
\eeq
Although in one dimension such order parameters with continuous symmetry do not have vacuum averages, at $T=0$ their correlation functions exhibit slow (algebraic) decay. In the present case 
\beq
<\Psi_+(x,0)\Psi^\dagger_+(x',0)> \sim |x-x'|^{-2K_+}.
\eeq
However, there may be operators with correlators decaying faster than that of $\Psi_+$. They are different in different phases of our model.  Depending on which of the cosines in (\ref{charge2}) takes over, the ground state of this model describes either quasi long range superfluid order in each tube or pair density wave state.  The latter state has a singularity in the density-density correlation function at the finite wave vector $2k_F$. 

  When both cosines are relevant (that is at $1/2 < K_- <2$) these states are separated by a quantum critical point (QCP), the location of which is approximately determined by the relation
  \beq
(\tilde t_{\perp}/\Lambda)^{K_-} \sim (\tilde V_1/\Lambda)^{1/K_-},
\eeq 
 where $\Lambda$ is the ultraviolet cut-off determined by the lattice. 
 The vicinity of the QCP can be studied analytically when $K _-\approx 1$ (this will be our assumption throughout the rest of the paper). The result for $K_-=1$ is that the transition belongs to the Ising model universality class. By continuity this continue to hold throughout the entire region of existence of QCP $1/2 < K_- <2$.

 For $K_- \approx 1$ it is convenient to refermionize (\ref{charge2}) with the result 
\begin{widetext}
  \beqa
   H_- =
  \int \rd x \Big\{ \frac{\ri v_-}{2}(-\rho_R\p_x\rho_R + \rho_L\p_x\rho_L - \eta_R\p_x\eta_R + \eta_L\p_x\eta_L) + \nonumber\\
   4\pi v_-(K_- -1)\rho_R\rho_L\eta_R\eta_L +  2\ri m_+\rho_R\rho_L + 2\ri m_-\eta_R\eta_L\Big\}, 
 \label{fin}
 \eeqa
\end{widetext}
Where $m_{\pm} = \tilde t_{\perp} \pm \tilde V_1$ and $\rho_{L,R}$ and $\eta_{L,R}$ are left- and right-moving components of Majorana (real) fermions:
\beqa
&& \rho_{R,L} = \frac{1}{\sqrt{2\pi a_0}}\cos\Big[\sqrt\pi(\Phi_-\pm\Theta_-)\Big], \nonumber\\
&& \eta_{R,L} = \frac{1}{\sqrt{2\pi a_0}}\sin\Big[\sqrt\pi(\Phi_-\pm\Theta_-)\Big].
\eeqa
We comment that the fermionization of the system in terms of real (Majorana) fermions is consistent with the Ising type symmetry breaking. 
This model is equivalent to the continuum limit of two quantum Ising (QI) models coupled by the energy density operators \cite{Tsvelik_2011}. The transition occurs when one of the Majorana masses becomes zero. To access the correlation functions we need  to express the original spin operators in terms of the Ising model fields:
\begin{widetext}
\beqa
 \sigma_a^+ = \re^{\ri\sqrt\pi\Phi_+} \Big(s_+s_- \pm \ri \mu_+ \mu_- \Big) + ... ;  
\sigma^z_a = \frac{1}{\sqrt{2\pi}}\p_x\Theta_+ + C^z\Big[\re^{\ri Qx +\ri\sqrt\pi\Theta_+}\Big(s_+\mu_- \pm \ri \mu_+ s_-\Big) + h.c.\Big)\Big] +... \label{OPs}
\eeqa
\end{widetext}
where $Q = \pi/a_0 + 2k_F$, $s_{\pm}$ ($\mu_{\pm}$) are Ising order (disorder) parameters for the Ising models represented by $\rho$ and $\eta$ fermions respectively. These operators are nonlocal in terms of fermioms. Their explicit expressions are not needed here, all we need to know is that 
 in the part of the phase diagram  $m>0 $ we have $\la \sigma\ra \neq 0, ~~ \la \mu\ra =0$ and for $m<0$ we have  $\la\sigma\ra = 0, ~~ \la\mu\ra \neq 0$. Therefore at $\tilde t_{\perp} > \tilde V_1 >0$ when both masses have the same sign both $\sigma_{\pm}$ have vacuum averages. Replacing these operators in (\ref{OPs}) by  their vacuum averages we get  
\beqa
\sigma_a^+ \sim \re^{\ri\sqrt\pi\Phi_+}[<s>]^2.
\eeqa
Since $\Phi_+$ has a gapless spectrum, the corresponding correlation function decays algebraically with the exponent $K_+/2$ which is four times smaller than the exponent for $\Psi_+$ (\ref{psi2}). In the other  phase $m_{+}m_- <0$  and a similar replacement can be done for the oscillatory part of the density operator yielding 
\beqa
\sigma^z_a \sim \pm \cos(Qx + \sqrt{\pi}\Theta_+)[<s>_+<\mu>_-],
\eeqa
so that 
\beq
< \sigma^z_a(x,0)\sigma^z_b(x',0)> \sim \frac{\cos[Q(x-x')]}{|x-x'|^{1/2K_+}}.
\eeq
The latter situation corresponds to Pair Density Wave (PDW). The density oscillations in this phase exist alongside the superfluidity of coupled pairs, though in the presence of disorder PDW is pinned \cite{shimshoni}  and the superfluidity is not destroyed. The QCP separating the two phases occurs when one of the masses (if $t_{\perp} >0, V_1 >0$ it is always $m_-$) becomes zero. 
Thus, for two chains $N=2$ the transition occurs in one Ising model and since Matsubara time correlation functions of quantum Ising model  at $T=0$ are the same as the correlation functions of 2D classical Ising model, it belongs to  the universality class of $d=2$ classical Ising model. The correlation length exponent is $\nu=1$.

\subsection{Correlation Functions}

The SF-CSF transition can be determined from measurements of various correlation functions. Here we will specifically address the situation when
the dipoles are indirect excitons confined in bi-layered structures. Then, since such excitons can be converted directly into light, 
 the excitation spectrum can be extracted from  measurements of light emission. Such emission is described by the linear coupling of the electric field $\bf E$ to
the excitonic operators $  \hat a_{z,x}, \hat a^+_{z,x}$:
\beqa
H_{int} =- \sum_{x,z} [({\bf E d})\hat a_{z,x} + ({\bf E^* d})\hat a^+_{z,x}]
\label{Eint}
\eeqa
Therefore in the first order of perturbation theory in the excitonic transition matrix element $d$ the emission or absorbtion probability of a single photon is related to the imaginary part of the $<\sigma^+\sigma^->$ correlation function. According to (\ref{OPs}) we have 
\beqa
&& \Sigma = <\hat T\sigma_1^+(x,t)\sigma_1^-(0,0)> + <\hat T\sigma_2^+(x,t)\sigma_2^-(0,0)>= <\hat T\exp(i\sqrt{\pi}\Phi(t,x)\exp(-i\sqrt\pi\Phi(0,0)> \nonumber\\
&& \cdot [<\hat T(s_+s_-)(x,t)
(s_+s_-)(0,0)> + <\hat T(\mu_+\mu_-)(x,t)(\mu_+\mu_-)(0,0)>] = \nonumber\\
&& <\hat T\exp(i\sqrt{\pi}\Phi(t,x)\exp(-i\sqrt\pi\Phi(0,0)>[G_{s_+}(x,t)G_{s_-}(x,t) + G_{\mu_+}(x,t)G_{\mu_-}(x,t)], \label{corrfun}
\eeqa
where $G_{\sigma}$ and $G_{\mu}$ are two-point correlation functions of $s$ and $\mu$ operators. These correlators of the Ising model are well known. In the ordered state of the Ising model the Lehmann expansion for $G_{s}$ contains matrix elements between the vacuum and the states with even number of Majorana fermions (including zero) and for $G_{\mu}$ it contains matrix elements with the odd number of fermions \cite{berg}. In the disordered state $s$ and $\mu$ are interchanged. Keeping this in mind in the SF phase we obtain the following expansion for (\ref{corrfun}):
\beqa
&& \Sigma(\tau,x)= \frac{(<s_+><s_->)}{[\tau^2 + (x/v)^2]^{\gamma}}\times \label{2point}\\
&& \Big\{1 + \frac{1}{(2\pi)^2}\int d\theta_1d\theta_2 [\tanh(\theta_{12}/2)]^2\exp\Big[- m_-|\tau|\Big(\cosh\theta_1 + \cosh\theta_2\Big) + im_-(x/v_-)\Big(\sinh\theta_1 + \sinh\theta_2\Big)\Big] + ...\Big\}  \nonumber
\eeqa
where $ \theta_{12} = \theta_1 - \theta_2, \gamma = K_+/4$ and the dots stand for the matrix elements between the vacuum and states with  energies higher than $2m_-$.

\begin{figure}
\centerline{\includegraphics[angle = 0,
width=0.5\columnwidth]{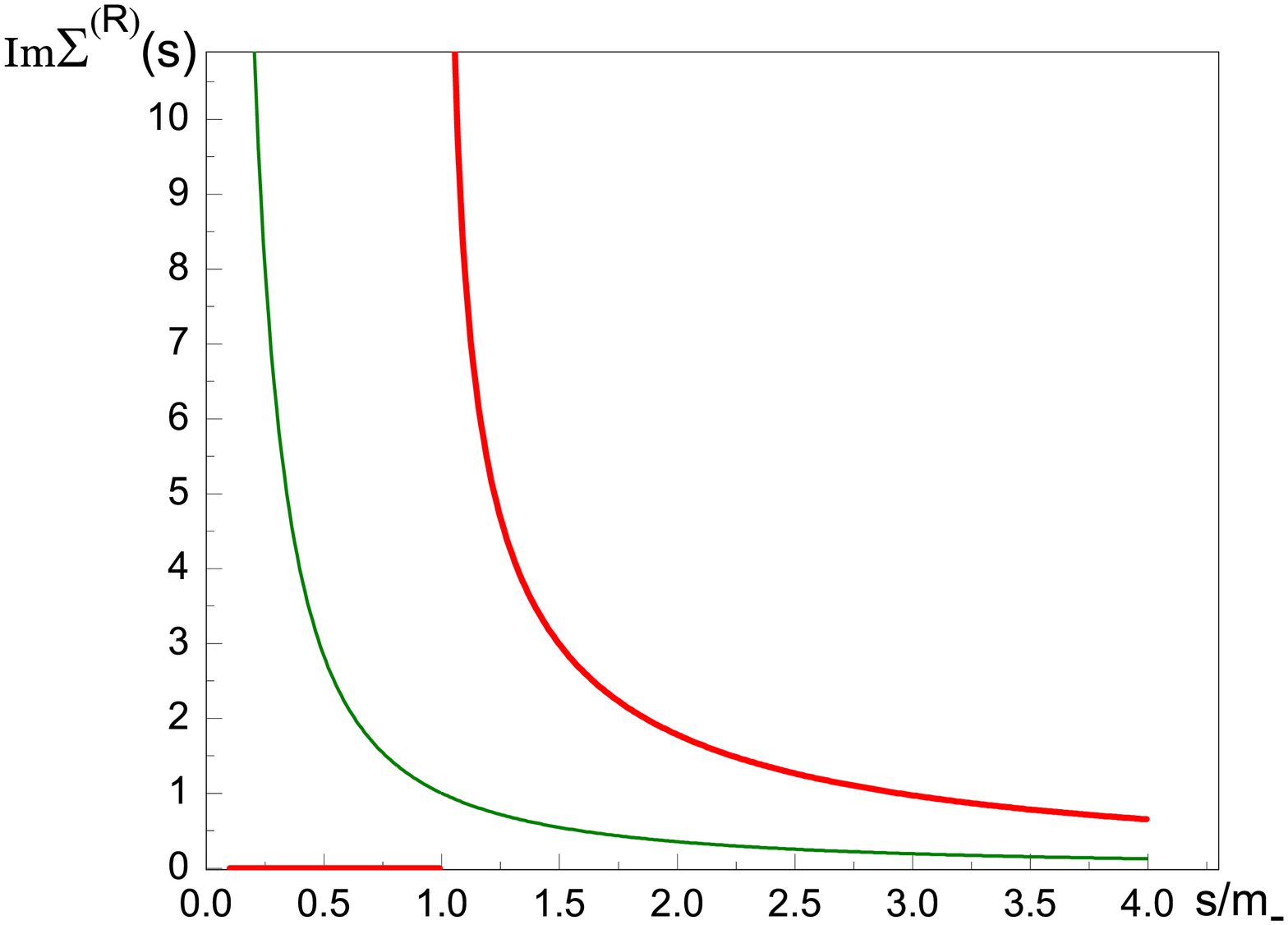}} 
\caption{(Color online)
 Single photon emission intensity given by Eqs.(\ref{abs},\ref{26}) for the case $K_+=1$. The thin green line corresponds to the emission from the SF phase and the thick red line describes the intensity from the CSF phase.}  \label{sig}
\end{figure}

For simplicity in what follows  we will set $v_+ = v_-$. Performing the Fourier transformation of (\ref{2point}) and the analytic continuation from Matsubara frequency to real one $i\omega \rightarrow \omega + i0$ and then taking the imaginary part, we obtain 
\beqa
\frac{\Im m\Sigma^{(R)}(\omega,k)}{<s_+><s_->} = \frac{Z}{\Big[\omega^2 - (vk)^2 \Big]^{1-\gamma}} \theta(|\omega| - v|k|)+ \Sigma^{(2)}(\omega,k)\label{abs}, 
\eeqa
where $Z \equiv 2^{1-2\gamma}\sin[\pi(1-\gamma)]\Gamma(1-\gamma)/\Gamma(\gamma)$ and
\beqa
&& \Sigma^{(2)}(s^2) = \int^{\cosh^2\theta = s^2/4m_-^2}_0 d\theta [\tanh\theta]^2\Big(\frac{s^2}{4m_-^2\cosh^2\theta}-1\Big)^{2\gamma-1}F\Big(\gamma,\gamma,2\gamma;1- \frac{s^2}{4m_-^2\cosh^2\theta}\Big) =\nonumber\\
&& \int^{1- (2m_-/s)^2}_0 \frac{dx x^2}{1-x^2}\Big(\frac{s^2}{4m_-^2}(1-x^2)-1\Big)^{2\gamma-1}F\Big(\gamma,\gamma,2\gamma;1- (s/2m_-)^2  +(s/2m_-)^2x^2\Big),\label{cont}
\eeqa
with $s^2 = \omega^2 - (vk)^2$. The thin green line features the dependence (\ref{abs}) in Fig.~\ref{sig}.
In the vicinity of  the 2-particle threshold $s^2 = 4{m_-}^2$ (\ref{cont})  behaves as 
\beqa
\propto (s^2/4m_-^2-1)^{2+2\gamma}
\eeqa
indicating that the 2-particle continuum is very weak. Thus in the superfluid  phase the absoption is dominated by the first term in (\ref{abs}) originating from the gapless excitations of the condensate. 

 To observe clear signs of the  Majorana mode one has to make measurements in Pair Density Wave phase where the gapless excitations can be emitted only together with one Majorana fermion. The dominant contribution comes from $G_{s+} G_{s-}$ term in (\ref{corrfun}); the function $G_{s+}$ is replaced by constant as before, but $G_{s-}$ now contains the emission of one Majorana fermion. As a result  we get 
\beqa
 \Sigma(\tau,x) = 
 \frac{(m_+|m_-|)^{1/4}}{[\tau^2 + (x/v)^2]^{\gamma}}\Big\{\frac{1}{(2\pi)}\int d\theta \exp\Big[- |m_-||\tau|\cosh\theta  + im_-(x/v_-)\sinh\theta\Big] + ...\Big\}  
\label{25} 
\eeqa
where the dots stand for the terms containing emissions of at least three massive particles. Therefore in the interval $(3m_-^2 > \omega^2 - (vk)^2 > m_-^2$) we have: 
\beqa
 \frac{\Im m\Sigma^{(R)}(s^2)}{<s_+><s_->}  = 
Z  \frac{[\Gamma(\gamma)]^2}{2\Gamma(2\gamma)}s^{-2\gamma}\Big(s^2 - m_-^2\Big)^{2\gamma-1}F\Big(\gamma,\gamma,2\gamma; 1- m_-^2/s^2\Big)\theta(s^2 -m_-^2).
\label{26}
\eeqa 
 As we see, the spectral function here has a strong singularity at the one-particle threshold if $2\gamma -1 <0$ (that is, $K_+ <2$). Such feature is shown by the thick red line in Fig.~\ref{sig}. 



\subsection{ N tubes symmetrically coupled}

Now we consider a system of $N >2$ tubes coupled to each other in such a way that each tube interacts with all others. The treatment in this case is more complicated since we have to resort to non-Abelian bosonization (see, for instance, \cite{boson},\cite{dms}). As a starting point we take non-interacting tubes with SU(2) symmetry ($V_0 = t_{||}$). Then an invidual  tube is equivalent to spin S=1/2 isotropic Heisenberg antiferromagnet and in the continuum limit  is described by the SU$_1$(2) Wess-Zumino-Novikov-Witten (WZNW) model. The sum of $N$ SU$_1$(N) WZNW Hamiltonians can be decomposed as (see, for instance \cite{nachlich})
\beqa
SU_1(2) + ... SU_1(2) = \Big[SU_2(N)/U^{N-1}(1)\Big]\times Z_N\times U(1) \label{decom}
\eeqa
This decomposition should be understood in the sense that operators (primary fields) of the critical theory on the left hand side of the identity can be written as products of operators belonging to the critical field theories on the right hand side. Decompositions  of that kind can be very helful outside criticality if the perturbation happens to be such that it does not act in at least one of the sectors. Decomposition (\ref{decom}) is consistent with the fact that central charges of the theories on the left- and right hand side of (\ref{decom}) are equal:
\beqa
N = \Big[\frac{2(N^2-1)}{N+2} - (N-1)\Big] + \Big(\frac{3N}{N+2} -1\Big) + 1.
\eeqa  

The U(1) subsector of (\ref{decom}) corresponds to the symmetric bosonic phase (N-tube generalization of $\Phi_+$ from the previous subsection). Since the inter-tube interaction does not contain this field, it remains gapless. As far as the other sectors are concerned, we will leave a detailed analysis to future publications and only formulate some conjectures. Information extracted from our numerical calculations suggests the following scenario. Close to the critical point the $\Big[SU_2(N)/U^{N-1}(1)\Big]$-sector (Gepner's parafermions \cite{Gepner}) is weakly coupled to the rest. This coset sector remains massive throughout the entire phase diagram, at least in the part where $t_{\perp} >0, V_1 >0$. Its analog for $N=2$ is $\rho$ Majorana fermion. The Z$_N$ sector is the one where the critical point is located. The relevant perturbation around the critical point can be guessed from the numerics which yields $\nu = 5/6\approx 0.833$ for $N=3$ and $\nu = 0.75$ for $N=4$. Using the relation $\nu = 1/(2-d)$, where $d$ is scaling dimension of the operator responsible for the deviation from criticality, we find $d \approx 0.8$ for $N=3$ and $d \approx 0.67$ for $N=4$. On the other hand in the model of Z$_N$ parafermions there is an operator with scaling dimension $d = 4/(N+2)$\cite{ZF,Gepner} which reproduces perfectly the numerical values of $d$ (see below).

\section{Numerical results}\label{numerics}
The QPT transition discussed above is characterized by {\it disappearance} of the algebraic off-diagonal order in all M-body density matrices, where $M=1,2,...,N-1$.
Specifically, as the inter-tube interaction is increasing the order existing in {\it all} M-body density matrices must eventually vanish up to the order $M=N-1$.
At the same time, the  order remains, practically, unaffected in the N-body density matrix.

\subsection{M-body density matrix}
The $M$-body density matrix $D_M$ can be written explicitly as
\beq
D_M(\left\{(x_1,z_1),...,(x_M,z_M)\right\};\left\{(x'_1,z'_1),...,(x'_M,z'_M)\right\})  =\langle \prod_{m=1,...,M}a^\dagger_{x_mz_m} \prod_{m'=1,...,M}a_{x_{m'}z_{m'}}\rangle 
\label{D_M}
\eeq 
where $\langle ... \rangle$ stands for the quantum-thermal averaging. 

In 1D SF $D_1(x,z; x',z')\sim 1/|x-x'|^b, \,\, b<1,$ exhibits algebraic order at large
$|x-x'|$. In the CSF, $D_1(x,z; x',z')\sim \exp(-|x-x'|/\xi_0),\, \xi_0 \sim 1$, that is, it becomes short ranged at $T=0$ regardless of the filling factor.
Thus, in the CSF phase the $N$-body density matrix is characterized by the exponential decay
$D_N(x_1,...,x_m;x'_1,...,x'_m) \sim \exp(-|x_{m_1}-x_{m_2}|/\xi_0)$ with respect to any pair $x_{m_1},\, x_{m_2}$ of coordinates from the set  $x_1,...,x_m$ (or   $x'_1,...,x'_m$). In the CSF there is also the algebraic order $D_N \sim 1/|R_{cm}-R'_{cm}|^c,\, c>0,$ with respect to the "center of mass" coordinates
 $R_{cm} = [x_1 + ... + x_N]/N$ and $R'_{cm} = [x'_1 + ... + x'_N]/N$ defined, respectively, for the first and the second sets of the coordinates as introduced in Eq.(\ref{D_M}), provided $|R_{cm} -x_m| \leq \xi_0$ and $|R'_{cm} -x'_m| \leq \xi_0$ for all $m$.

The transition SF to CSF can be detected by the critical behavior of any density matrix. In particular, the same criticality controls the long-distance behavior of $D_N$  with respect to $|R_{cm} -x_m|$ (or  $|R'_{cm} -x'_m|$). That is, in SF phase $D_N$ is trivially long-ranged with respect to $|R_{cm} -x_m|$ because $D_N$ can simply be factorized into a product of $D_1$. In contrast, in the CSF-phase, while exhibiting ODLRO with respect to $R_{cm}-R'_{cm}$, $D_N$ is short-ranged with respect to $|R_{cm} -x_m|$ (or  $|R'_{cm} -x'_m|$). Thus, the criticality can also be detected by measuring the behavior of the relative distances $x_m$ (or $x'_m$). Specifically, we have considered 
the square of so called {\it gyration radius} \cite{JLTP} as the mean of 
\beq
R^2_g=\frac{1}{N^2}\sum_{m, n=1,2,...,N}\left[ x_m - x_n\right]^2
\label{R^2}
\eeq
 with respect to the first set of the coordinates of $D_N$ defined in Eq.(\ref{D_M}), provided the coordinates from the second set are kept within some distance $\sim \xi_0 $ from $R'_{cm}$ \cite{note2}. More specifically, $x_k$, where $k=1,2,..,N$, represents the x-coordinate in the $k-$th tube.
  
In the SF of a length $L$, $R^2_g=R^2_{0} \approx \frac{1-b}{4(3-b)}L^2 \sim {\cal O}(L^2) $, and in the CSF $R^2_{SCF} \sim {\cal O}(1) \approx \xi^2_0 << L^2$. In what follows we will be calculating the mean of the ratio $G_N=R^2_g/R^2_0$, so that it is changing from $G_N\approx 1$ in the SF state to $G_N\sim 1/L^2 \approx 0$ in the CSF phase.
It is worth mentioning that $R_g$ can be viewed as a typical width of a chain. For strongly bound case this width is $\sim \xi_0 \approx 1$, and in the SF phase it is $\sim L$, and, thus, it exhibits critical behavior typical for correlation length.  
\begin{figure}
\centerline{\includegraphics[angle = 0,
width=0.5\columnwidth]{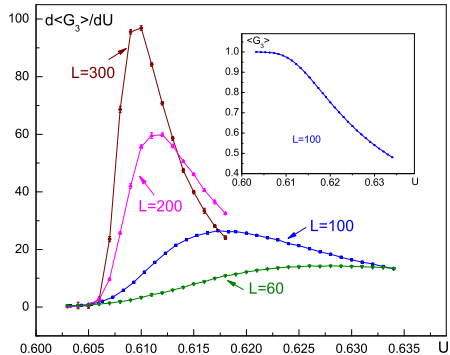}} 
\vspace{-.70cm}
\caption{(Color online)
$d\langle G_3\rangle /dU$ versus the interaction strength $U$ for $L=60,100,200,300$ with $\beta=L$. Inset: $\langle G_3\rangle$ versus $U$ for $L=100$. The SF phase corresponds to $\langle G_3\rangle\approx 1$ and the CSF to $\langle G_3\rangle\approx 0$.
The transition point SF-CSF for a given size can be identified by the maximum of $dG_3/dU$ reaching the thermodynamics limit at $U_c \approx 0.61$. 
}\label{L100}
\end{figure}
 \begin{figure}
\centerline{\includegraphics[angle = 0,
width=0.5\columnwidth]{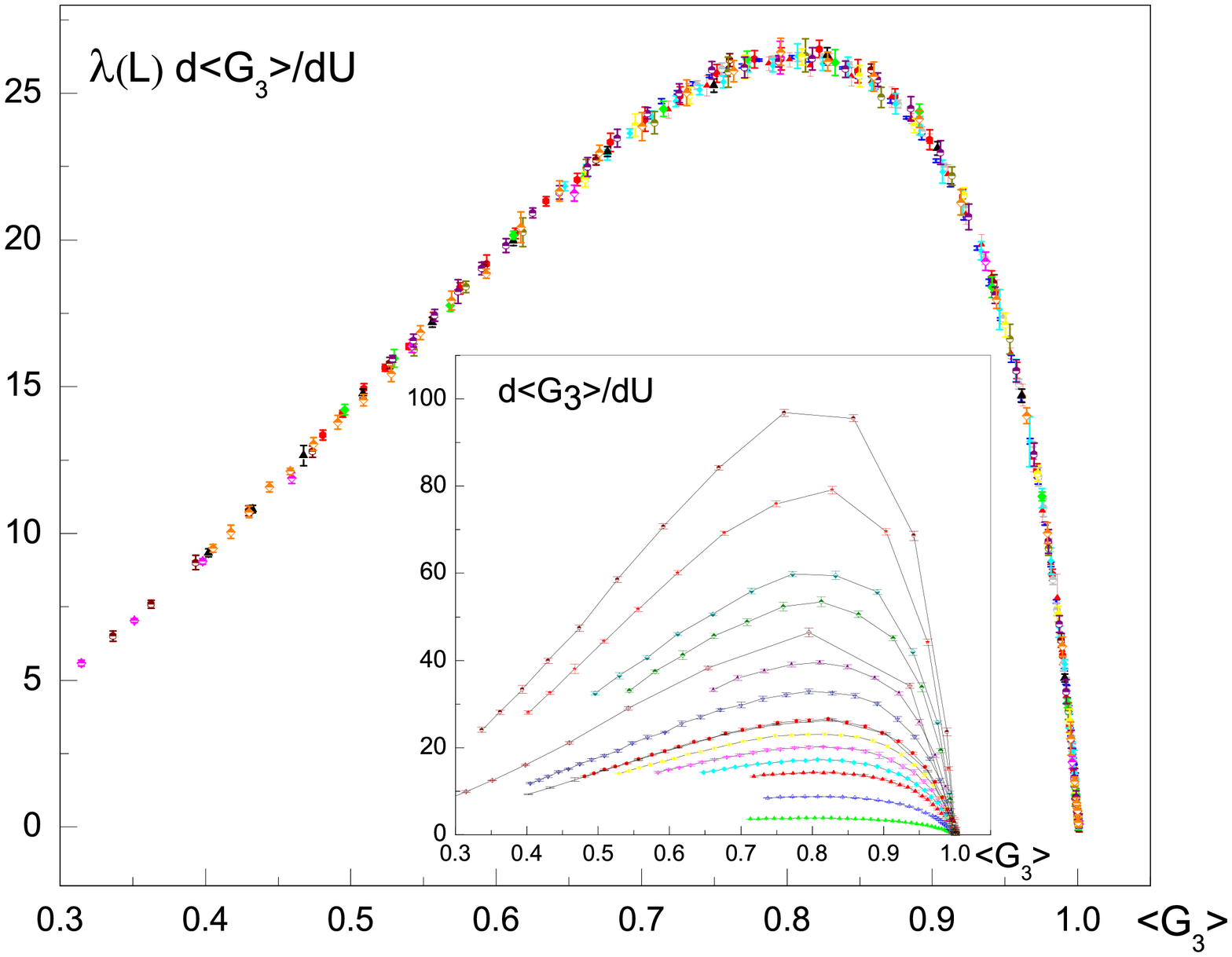}} 
\vspace{-.70cm}
\caption{(Color online)
$d\langle G_3\rangle /dU$ versus $\langle G_3\rangle$ for sizes $L=20,40, ..., 300$ rescaled by a factor $\lambda(L)$ in order to achieve collapse to the curve $L=100$ ($\lambda(100)=1$). Inset: $d\langle G_3\rangle/dU$ versus $\langle G_3\rangle$ for the same sizes.
}\label{G3}
\end{figure}
\begin{figure}
\centerline{\includegraphics[angle = 0,
width=0.5\columnwidth]{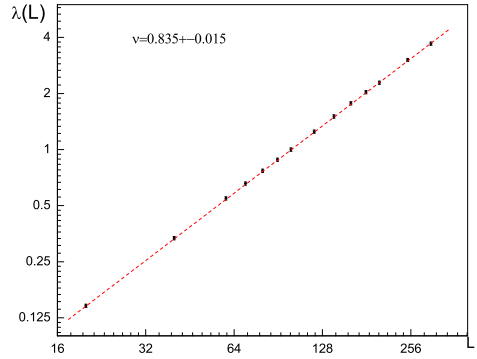}} 
\vspace{-.70cm}
\caption{(Color online)
The rescaling factor $\lambda^{-1} (L)$ versus $L$ for $N=3$ from Fig.\ref{G3}. The slope gives the correlation length exponent $\nu =0.835 \pm 0.015$. 
}\label{nu3}
\end{figure}
\begin{figure}
\centerline{\includegraphics[angle = 0,
width=0.5\columnwidth]{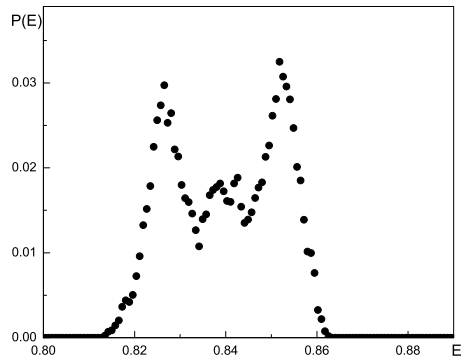}} 
\vspace{-.70cm}
\caption{(Color online)
Energy histogram $P(E)$ for $N=5$ tubes with $L=400, \, \beta=400$. The first-order transition is determined by the value of $U=U_c,\, U_c=0.7235$,
corresponding to the situation when the histogram becomes bimodal. 
}\label{Iorder}
\end{figure}
\begin{figure}
\centerline{\includegraphics[angle = 0,
width=0.5\columnwidth]{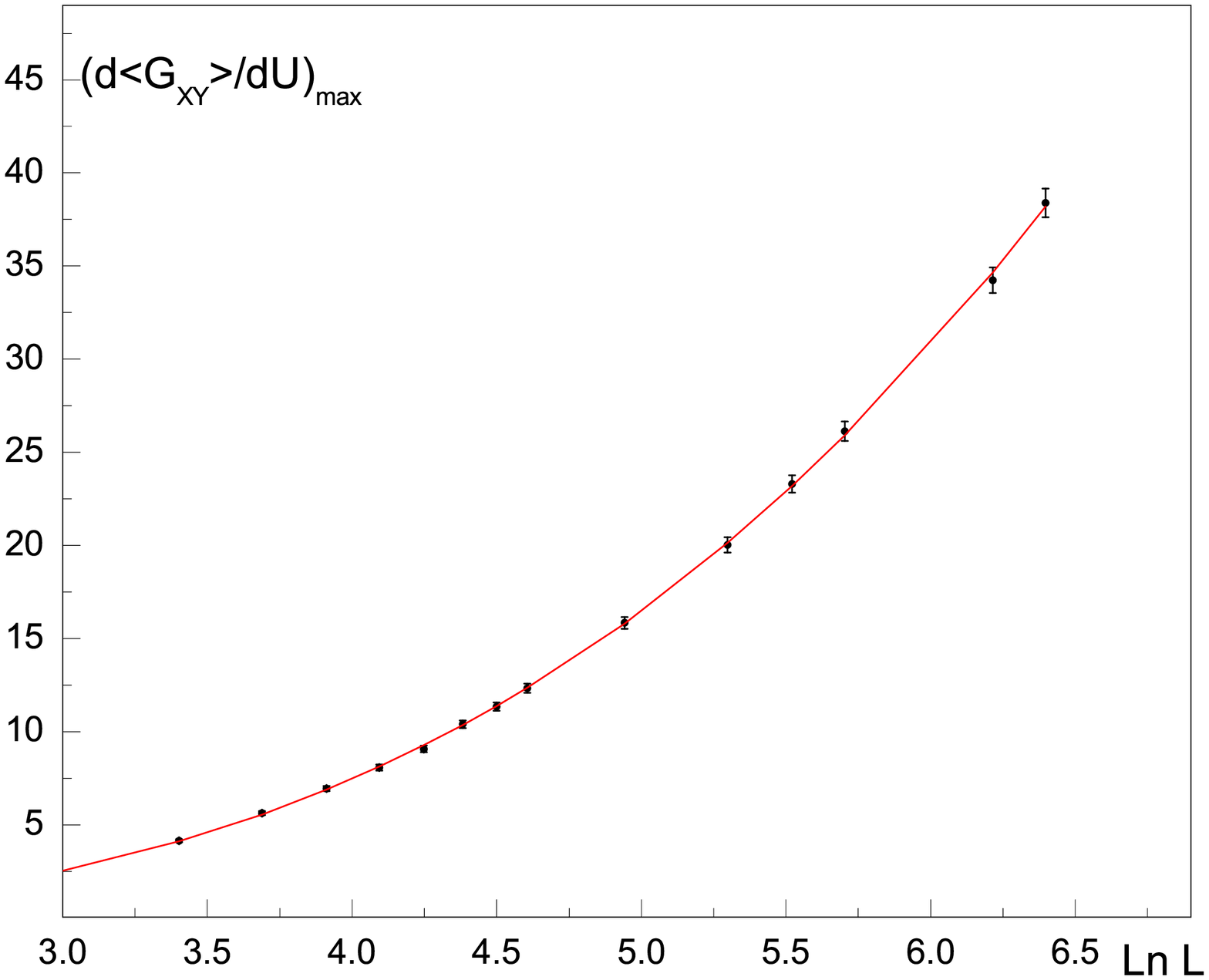}} 
\vspace{-.70cm}
\caption{(Color online)
The maximum value of $d\langle G_{XY} \rangle /dU$ versus $\ln L$ for the case of zero inter-tube tunneling, $N=4$, $K(\hat{z})=\infty$. The solid red line is the fit by the
finite size scaling ansatz for the BKT-transition: $d\langle G_{XY} \rangle/dU = A \ln^3(L/L_o), \, A=0.205,\, L_o=1.98$. 
}\label{XY}
\end{figure}
\subsection{J-current formulation}\label{J-Hamilton}
Hamiltonian (\ref{Habi}) and the gyration radius (\ref{R^2}) have been used for {\it ab initio} simulations of a single chain (with exactly one polar particle per layer (in $d=2$) or tube (in $d=1$)  for the case $t_\perp=0$ \cite{JLTP}. It was found that the chain can undergo quantum roughening transition with the tuning parameter being the interaction strength $V_d$.
The transition is, practically, insensitive to the interaction range.  

The Monte Carlo simulations at finite densities $n <1$ ( incommensurate with the lattice along the tubes)  in each layer have been conducted in the discrete-time $J$-current-type formulation \cite{Jcurrent} of the Hamiltonian (\ref{Habi}) \cite{JLTP}. For the purpose of analyzing the universality of the transition this approach turns out to be much more efficient than the {\it ab initio} one.
Here we will be using the model where the inter-tube tunneling is allowed. The actual dipole-dipole interaction will be replaced by onsite attraction between neighboring tubes, with the intra-tube dipole-dipole repulsion ignored. This approximation becomes essentially exact when $n << 1$: while the inter-tube attraction is not affected, the intra-tube dipole-dipole repulsion between atoms scales as $ \sim n^3 \to 0$ in 1d and, thus, becomes irrelevant. The corresponding space-time action, then, becomes
\beq
H_J = \sum_{b}\left[ \frac{K(\hat{b}) (\vec{J}_{b})^{2}}{2} +\frac{U(\hat{b})(\nabla_z \vec{J}_{b})^2}{2} - \mu J^{(\hat{\tau})}_{b}\right], %
\label{HJ}
\eeq 
where $\vec{J}_{b}$ is the integer bond current obeying Kirchhoff's conservation law \cite{Jcurrent}. In some sense,
these conserved currents represent world-lines of particles in imaginary discrete space-time, with $H_J$ being the action in Feynman's path integral.
The summation in (\ref{HJ}) is performed over all space-time bonds $b$ (coming out from a space-time site ($x,\tau, z$) either along $\pm \hat{x}$ or along imaginary time $\pm \hat{\tau}$ or along $\pm \hat{z}$ directions);  $\nabla_{z} \vec{J}_b \equiv \vec{J}_{b}(x,\tau,z+1) - \vec{J}_{b}(x,\tau,z)$; $\mu$ denotes chemical potential. [Here we tuned $\mu$ to have 1/2 filling of bosons per site in each tube]. Periodic boundary conditions along space $0<x<L-1, 0\leq z \leq N-1$ and along imaginary time $ 0\leq \tau \leq \beta$, with $\beta=L$, where $L=2,3, ....$, have been used. 
The coefficients $K,U$ can be related to $t_{z,z'}$ and $V_{xz;x'z'}$ from Eq.(\ref{Habi}): $K(\hat{z}) \sim 1/t_\perp$, $K(\hat{x})=K(\hat{\tau}) \sim 1/t_{||}$, 
$U\sim V_d$. The case $K(\hat{z})=\infty$ corresponds to zero inter-tube tunneling (studied in Ref.\cite{JLTP}). Here we will focus on $ K(\hat{z})= K(\hat{x})=K(\hat{\tau})$ situation as the one which naturally represents the whole universality class. 

It is worth emphasizing that the inter-tube attraction between two particles (in neigboring tubes) located, respectively, at the space-time points $(x,z,\tau)$ and $(x,z\pm 1,\tau)$ is described by the terms $ \sim  - U \vec{J}(x,z,\tau) \vec{J}(x,z\pm 1,\tau)$. Accordingly, when these two particles form a bound state, their world-lines stay close to each other to gain the binding energy $\sim U$. Similarly, a chain of several particles is represented by a bundle of several world-lines forming a membrane in the space-time.

The action (\ref{HJ}) can be viewed as a coarse grained dual representation
of the Hamiltonian (\ref{Habi}). While being not precise for quantifying finite energy (non-universal) properties of the system, the J-current model \cite{Jcurrent} belongs to the same universality class as the original model (\ref{Habi}). Thus, for the purpose of this work and for sake of numerical practicality, it will be sufficient to study the model (\ref{HJ}). 

We also note that, while being formally defined on the lattice, the model (\ref{Habi}) and its dual formulation (\ref{HJ}) account well for the
continuous space-time situations at low energies as long as the filling factor $n$ remains incommensurate with the lattice. Long-range intra-tube repulsion may complicate the situation by inducing crystalization at, say, $n=0.5$ and, thus, shifting the CSF phase to lower densities. Such feature, however, does not affect the
universality of the N-SF to CSF transition, and, in order to establish it in a most efficient way we simply turn off the intra-tube repulsion and study the case $n=0.5$.

Monte-Carlo simulations of the model (\ref{HJ}) have been performed within the Worm Algorithm approach \cite{WA}. 
 Green's function in imaginary time (as well as the density matrix (\ref{D_M})) is given by the statistics $D_N(\{x_m,\tau_m,z_m\};\{x'_m,\tau'_m,z'_m\})$
  of "sources" and "sinks"
of the bond currents located, respectively, at $(x_m,\tau_m,z_m),\, m=1,2,...,N$, and $(x'_m,\tau'_m,z'_m),\, m=1,2,...,N$, lattice points.
In order to insure the condition  $|R'_{cm} -x'_m| \leq \xi_0$, while  $(x_m,\tau_m,z_m),\, m=1,2,...,N$, are free to take any value,
we have convoluted $D_N(\{x_m,\tau_m,z_m\};\{x'_m,\tau'_m,z'_m\})$ with $P=\exp(-\sum_{m,n} [|x'_m-x'_n| +|\tau'_m-\tau'_n|]/\xi_0)$ as
$D_N(\{x_m,\tau_m,z_m\}; R'_{cm})=\int Dx'D\tau' Dz' D_N P \delta\left(R'_{cm} -\sum_m x'_m/N\right)$ , and, accordingly 
have evaluated the means of the normalized gyration radius $\langle G_N \rangle$ and of the center of mass distance $\langle |R_{cm}-R'_{cm}|\rangle$ where
 $\langle ... \rangle \equiv \tilde{Z}^{-1} \int Dx D\tau Dz dR'_{cm} ... D_N, \,\, \tilde{Z}= \int Dx D\tau Dz dR'_{cm} D_N$.

For sake of numerical efficiency we have symmetrized the model (\ref{HJ}) by choosing $U(\hat{b})$ independent of the type of a bond, that is, $U(\hat{b})=U$. 
The CSF phase has been identified by the condition $\langle |R_{cm}-R'_{cm}|\rangle/L =const $ and $\langle G_N\rangle \sim o(L^{-2})$ for $U>U_c$, where $U_c$ corresponds to the QCP.    
In the SF phase (that is, $U<U_c$), while the first condition remained, practically, unchanged, $\langle G_N\rangle \approx 1$ with high accuracy.
The criticality of the SF-CSF transition has been analyzed through evaluating the divergent behavior of $d\langle G_N\rangle /dU$ in the vicinity of $U=U_c$.

\subsection{Finite size scaling of the gyration radius}
As discussed above, $d\langle G_N\rangle/dU$ exhibits singularity in the limit $L \to \infty$. The change from $\langle G_N\rangle \approx 1$ to $\langle G_N\rangle \approx 0$ occurs in a narrow range $\delta U= |U-U_c|$ around the QCP $U_c \sim 1$. Such a behavior is clearly seen in Fig.~\ref{L100}:  the range of the transition $\delta U$ narrows as $L$ increases.  It is important to emphasize that in the following analysis of the criticality the knowledge of the exact value of $U_c$ in the thermodynamical limit is not required. All we need is an  approximate range where the derivative exhibits a clear sign of divergence.

The range $\delta U$ is controlled by the diverging correlation length $\xi(U) \sim |U-U_c|^{-\nu},\, \nu >0$. According to the finite size scaling approach, 
$\langle G_N\rangle$ can be represented as some regular function $F(y), \, y= L/\xi(U)$ varying from $F(y=0)=1$ to $F(y=\infty)=0$ over the range $ y \sim 1$.
Thus, $d\langle G_N\rangle/dU \approx F'y /\delta U \sim L^{1/\nu}$. 
Loosely speaking, one can view this relation as $d\langle G_N\rangle/dU \approx 1/\delta U, \,\, \delta U \approx L^{-1/\nu} \to 0$.  It should be noted that at any finite $L$ the actual divergence contains the so called {\it subleading} terms --  powers of $L$ smaller than $1/\nu$. These terms are the main source of systematic errors in our analysis.

 We have evaluated $d\langle G_N\rangle/dU$ numerically by Monte Carlo \cite{WA} and constructed 
the graphs $d\langle G_N\rangle/dU$ versus $\langle G_N\rangle$ by scanning over $U$ around the critical point $U_c$ for sizes $L=\beta = 10,20, ...300$. These graphs turn out to be self-similar so that
$d\langle G_N\rangle/dU$ for all sizes $> 10$ collapsed on a single master curve by simple rescaling of $ d\langle G_N\rangle/dU$ for size $L_1$ to another size $L_2$ as $ d\langle G_N\rangle/dU \to \lambda(L) d\langle G_N\rangle/dU$. Then, the rescaling coefficient $\lambda$, which represents the inverse width $\delta U$ as $\lambda \propto 1/\delta U$, has been plotted in the log-log$L$ axes in order to determine the critical exponent $\nu$.
The results of this procedure are presented on Figs.~\ref{G3},\ref{nu3} for the case $N=3$. The same procedure has been used in the cases $N=2,4$ as well. The found exponents are: $\nu=0.972 \pm 0.02$ for $N=2$, $\nu=0.835 \pm 0.015$ for $N=3$, $\nu =0.735 \pm 0.015$ for $N=4$. The errors include statistical errors as well as the systematic errors due to the subleading contributions. We note that the value of $\nu$ for $N=2$ is consistent with the $d=2$ Ising (or $q=2$ Potts) universality.  
We also note that the values of $\nu$ for $N=3$ and $N=4$ are consistent with the corresponding ones $\nu =0.837$ and $\nu = 0.756$ obtained by the Renormalization Group calculations for the $q=3,4$ 2d Potts model \cite{Dasgupta}. 

The above data collapse fails for $N>4$ after reaching some size $L \sim 50-100$ for $N=5$ and much smaller sizes for $N>5$. Furthermore, the energy histogram develops bimodality typical for I-st order transitions, Fig.~\ref{Iorder}. While for $N=5$ the bimodality develops on sizes $L\geq 400, \, \beta=L$, for $N=8$ it is  already  well developed  at $L=160, \, \beta=L$. Such features are consistent with the I-st order transition in 2d Potts model for $N=q>4$.

The finite size analysis has been applied to the case of zero inter-tube tunneling as well, when
the transition is expected to be in the BKT universality. That is, $\xi \sim \exp(...|U-Uc|^{-1/2})$. The variation of the
gyration radius $\langle G_{XY}\rangle $ in this case can also be represented by some regular function $F(y)$ characterized by the range $y\sim 1$ with $y=L/\xi$ . 
Thus, $d\langle G_{XY}\rangle /dU \approx F'y \sim (\delta U)^{-3/2}\sim (\ln(L/L_o))^3$  (where $L_o$ stands for some microscopic scale) at its maximum.
The maximum value of this derivative has been plotted as a function of $L=30,...,600 $ in Fig.\ref{XY}. As can be seen the fit of the data is consistent
with the $\ln^3L $ dependence with high accuracy.

\section{Summary and Discussion}
As shown above, the Majorana fermions description of the Ising-type transition to CSF state in the two-chain system is consistent with the numerical evaluation of the correlation length exponent $\nu =0.972 \pm 0.02$ (versus its exact value $\nu=1$). 
The quantum transitions in the cases $N=3,4$ are characterized by  emerging enlarged symmetries: Z$_3$ for $N=3$ and Z$_4$ for $N=4$. The predicted values $\nu=5/6 \approx 0.833,\, N=3, $ and $ \nu=3/4=0.75,\, N=4,$ are matched well by the corresponding numerical ones $\nu=0.835 \pm 0.015$ for $N=3$ and $\nu =0.735 \pm 0.015$ for $N=4$. The symmetry enlargement occurs despite the short-range nature
of the tunneling between the tubes. In other words, the critical behavior proceeds as though the tunneling between all tubes is the same.
Such feature --- identical interaction between all elements --- is typical for the standard Potts model \cite{Wu} and should be contrasted with the p-clock model. 


The detection of the CSF order as well as the criticality to SF state can be based on measuring field-correlators. In the case of the layered structures supporting indirect excitons \cite{Yudson} this means analyzing the excitonic emission of light. 
 In the SF phase the emission intensity tests directly sound-like excitations of excitonic Luttinger liquid similarly to Eq.(\ref{abs}) for $N=2$ case.  
In the CSF phase, the one-photon emission is controlled by the gapped parafermionic modes and, therefore, acquires the threshold similar to the case described by
Eq.(\ref{26}) for $N=2$. Both features exhibit threshold singularities as shown in Fig.~\ref{sig}. Thus, the light emission can become a crucial tool for detecting parafermionic excitations.

It is also worth mentioning that, as the system enters the CSF phase, there should appear a special feature in the correlated $N$-photon emission. 
Since light is linearly coupled to the excitonic operator and in the CSF phase the algebraic order exists only in the product of $N$ excitonic  operators, Eq.(\ref{Psi}),
such order (entanglement) will be imprinted on $N$ emitted photons.  We will consider specific proposal for detecting such $N$-photon entanglement in greater detail elsewhere.

When this work was prepared for publication we learned about the preprint by Lecheminant and Nonne \cite{nonne} which results have a substantial overlap with ours.

 

\section{acknowledgements}
We are thankful to Philippe Lecheminant for useful comments.
ABK was supported by the National Science Foundation
under Grant No.PHY1005527 and by a grant of computer time from the CUNY HPCC under NSF Grants CNS-0855217 and CNS - 0958379. AMT acknowledges a support from US DOE under contract number DE-AC02-98 CH10886.


\begin{thebibliography}{99}
\bibitem{note2} Using $D_N$ instead of $D_1$ for detecting the criticality has purely numerical advantage --- the algorithm turns out to be much more efficient.
\end{thebibliography}
\end{document}